\documentclass[runningheads]{llncs}
\usepackage{graphicx}
\usepackage[misc]{ifsym}
\usepackage{lipsum}

\usepackage[utf8]{inputenc}

\usepackage[utf8]{inputenc} 
\usepackage[T1]{fontenc}    
\usepackage{hyperref}       
\usepackage{url}            
\usepackage{booktabs}       
\usepackage{amsfonts}       
\usepackage{amssymb}
\usepackage{amsmath}
\usepackage{nicefrac}       
\usepackage{microtype}      
\usepackage{multirow}
\usepackage{xcolor}
\usepackage{soul}
\usepackage{graphicx}

\newcommand\blfootnote[1]{%
  \begingroup
  \renewcommand\thefootnote{}\footnote{#1}%
  \addtocounter{footnote}{-1}%
  \endgroup
}
\begin{document}
\title{Deep Network Interpolation for Accelerated Parallel MR Image Reconstruction}
\titlerunning{Deep Network Interpolation for Accelerated pMRI}
%
\author{Chen Qin\inst{1,2}\textsuperscript{(\Letter)}\and
Jo Schlemper\inst{1,3} \and
Kerstin Hammernik\inst{1} \and 
Jinming Duan\inst{4} \and \\
Ronald M Summers\inst{5} \and
Daniel Rueckert\inst{1}}
\authorrunning{C. Qin et al.}
%
\institute{Biomedical Image Analysis Group, Department of Computing, Imperial College London, London, UK \\ \email{c.qin15@imperial.ac.uk}\and
Institute for Digital Communications, School of Engineering, University of Edinburgh, Edinburgh, UK
\and
Hyperfine Research Inc., Guilford, CT, USA
\and
School of Computer Science, University of Birmingham, Birmingham, UK 
\and
NIH Clinical Center, Bethesda, MD, USA
\blfootnote{
Presented at 2020 ISMRM Conference \& Exhibition (Abstract \#4958)}}

\maketitle              
\begin{abstract}
We present a deep network interpolation strategy for accelerated parallel MR image reconstruction. In particular, we examine the network interpolation in parameter space between a source model that is formulated in an unrolled scheme with L1 and SSIM losses and its counterpart that is trained with an adversarial loss. We show that by interpolating between the two different models of the same network structure, the new interpolated network can model a trade-off between perceptual quality and fidelity. 
\end{abstract}
\section{Introduction}
Deep neural networks have demonstrated their capabilities in reconstructing accelerated magnetic resonance (MR) image \cite{aggarwal2018modl,duan2019vs,hammernik2018learning,qin2019convolutional,qin2019k,schlemper2019data}. However, models trained with mean-squared-error (MSE) or L1 loss tend to reconstruct smooth images while models trained with adversarial loss can recover rich textures but with unrealistic artefacts. To balance between these two effects, we employ a simple yet effective deep network interpolation approach which manipulates linear interpolation in the parameter space of multiple neural networks. We evaluate our method on a public multi-coil knee dataset from the fastMRI challenge \cite{zbontar2018fastmri}. Our results indicate that the strategy can effectively balance between data fidelity and perceptual quality.

\section{Methods}
The proposed source model named sensitivity network (SN) extends from the Deep-POCSENSE proposed in \cite{schlemper2019data}. It embeds the iterative optimisation scheme in a learning setting, which employs an unrolled architecture consisting of neural network-based reconstruction blocks interleaved by data consistency (DC) layers. Specifically, the reconstruction block updates the estimate of the sensitivity weighted combined image, while DC is performed coil-wisely in k-space. In our work, the reconstruction block is modelled by a Down-Up network \cite{yu2019deep} which has two complex-valued input and output channels. The network was trained with L1 and SSIM loss between reference image $x_{ref}$ and the reconstruction $x_{rec}$:
\begin{equation}
    L_{SN}(x_{rec}, x_{ref})=1-SSIM(x_{rec}, x_{ref})+\lambda L_1(x_{rec}, x_{ref}).
\end{equation}

To recover rich textures and details, we additionally propose to reconstruct images via an adversarial loss, where a discriminator is employed to identify if an input image is a fully sampled image or a reconstructed one. Specifically, we use the least squares generative adversarial network (LSGAN) for training the discriminator and reconstruction network in an adversarial way, as well as combining that with $L_{SN}$ loss as a complementary metric. Then the network can be trained by minimising the following loss function:
\begin{equation}
    L_{SN-GAN}(x_{rec}, x_{ref})=\gamma L_{SN}(x_{rec}, x_{ref})+L_{lsgan}(m \odot x_{rec}, m \odot x_{ref}).
\end{equation}
Here $L_{lsgan}$ represents the LSGAN formulation and $\odot$ is the pixel-wise product. We also introduce a binary foreground mask $m$ to focus more on the texture of foreground regions.

However, we observed that models trained with $L_{SN}$ loss tend to generate smooth images with relatively high quantitative scores, while those trained with $L_{SN-GAN}$ loss can reconstruct images that contain better details and textures but with probably hallucinated artefacts. To balance between the quantitative and qualitative performances, we propose to interpolate the networks in the parameter space \cite{wang2019deep}. In detail, let $\{G; \theta\}$ denote the mapping function $G$ of the image reconstruction model parameterised by $\theta$. Assume $\{G^{SN};\theta_{SN}\}$ is the model trained with $L_{SN}$ loss and $\{G^{SN-GAN};\theta_{SN-GAN}\}$ is trained with $L_{SN-GAN}$ loss, and both of them share the same network structure. To achieve a continuous and smooth transition between effects of these two models, a linear interpolation of corresponding parameters is applied to derive a new interpolated model $\{G^{interpSN};\theta_{interpSN}\}$, where
\begin{equation}
    \theta_{interpSN}=(1-\alpha)\theta_{SN}+\alpha\theta_{SN-GAN},
\end{equation}
with $\alpha \in [0, 1]$ as the interpolation coefficient. The interpolation is performed on all layers of the networks, including weights and biases. Note that the deep interpolation can be readily extended for multiple models with the same network architecture.	

\begin{figure*}[!t]
\centering
\includegraphics[width=\linewidth]{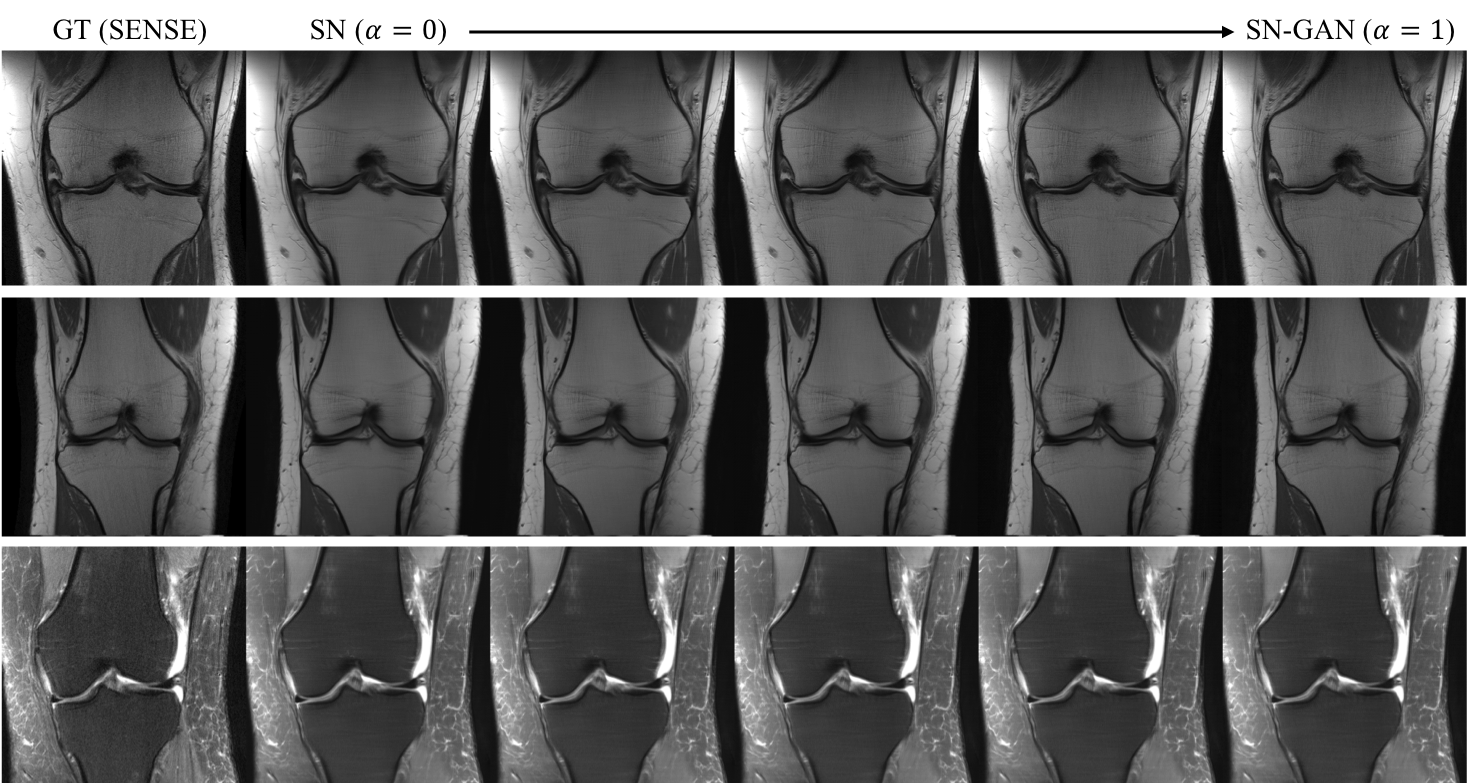}
\caption{Balancing the GAN and L1/SSIM effects with network interpolation for MRI reconstruction on acceleration factor 8. Source model (SN) generates smooth images, while SN-GAN recovers more details but with relatively low quantitative scores. By varying the interpolation coefficient $\alpha$ from 0 to 1, it allows for a smooth control of reconstruction effects, producing results that maintain both textures and fidelity.}
\label{fig1}
\end{figure*} 

\section{Experimental Settings}
Evaluation was performed on a public knee dataset provided by the fastMRI challenge \cite{zbontar2018fastmri}. The dataset contains 973 volumes for training and 199 volumes for validation, including both coronal proton-density weighting with (PDFS) and without (PD) fat suppression. The multi-coil data contains 15 channel array data, and we used a variable density Cartesian undersampling scheme with acceleration factor (AF) 4 and 8. In our experiments, both base model (SN) and GAN model (SN-GAN) were trained with a cascade number 10, and $\lambda=10^{-3}$ and $\gamma=0.1$ were chosen empirically. Specifically, the SN was first trained for 50 epochs using RMSProp with a learning rate $10^{-4}$, and then both models were further finetuned based on the pretrained SN model for 10 epochs with a learning rate $5\times 10^{-5}$. Here we use sensitivity encoding (SENSE) reconstruction \cite{uecker2014espirit} as ground truth, as it generates better images than root-sum-of-squares (RSS) reconstruction. 
\section{Results}
\begin{figure*}[!t]
\centering
\includegraphics[width=\linewidth]{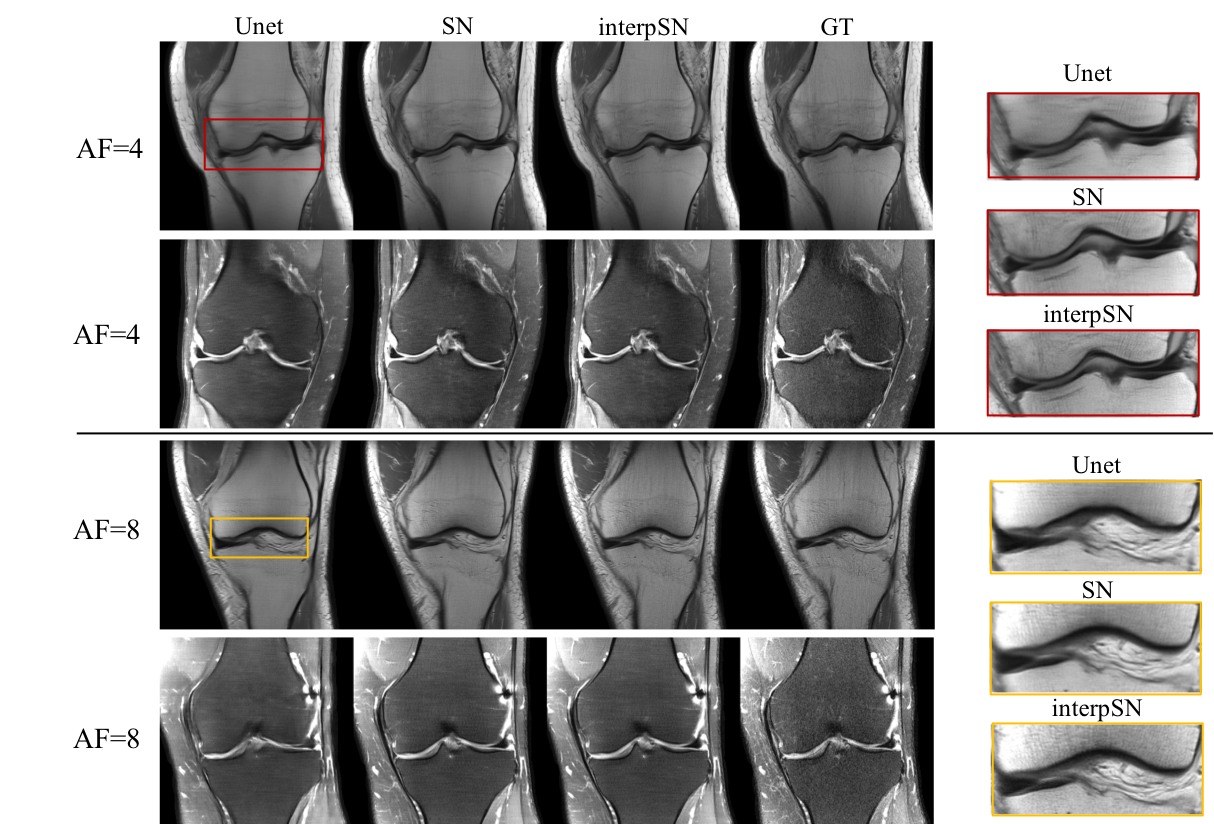}
\caption{Sample reconstructions on acceleration factor 4 and 8. The network interpolated model (interpSN) achieves much better results than the baseline model Unet which produces over-smooth images and does not guarantee consistence to the original $k$-space data. In comparison to Unet and SN, interpSN can generate visually sharper images and recover more details, especially on PD data. (Zoom in for better view)}
\label{fig2}
\end{figure*}

To examine the effect of the network interpolation, we present a group of qualitative results in Fig. \ref{fig1}, showing the visual quality changes of the reconstructed images by varying $\alpha$ from 0 to 1. Quantitative results are given in Table \ref{table1}, where interpSN stands for the model that interpolates between SN and SN-GAN models with $\alpha=0.5$. It can be seen that the interpolated model improves over its source model SN in terms of the textures and also outperforms SN-GAN in terms of quantitative scores. By adjusting $\alpha$, it can achieve a smooth transitions between the two effects without abrupt changes. Fig. \ref{fig2} also displays the sample reconstructions for each acquisition and AF respectively, and it shows that our model outperformed the baseline Unet \cite{zbontar2018fastmri} both quantitatively and qualitatively. Detailed visualisations indicate the capability of interpSN in recovering sharp textures over the other methods.

\begin{table}[!t]
    \centering
    \caption{Quantitative results on the fastMRI validation set for each acquisition and AF. The interpolated model (interpSN) is compared with our source model SN and SN-GAN model, as well as a baseline approach Unet \cite{zbontar2018fastmri}, in terms of normalised mean square error (NMSE), peak to-noise-ratio (PSNR) and structural similarity (SSIM). The fully sampled SENSE reconstruction \cite{uecker2014espirit} was used as ground truth, and all the metrics were evaluated only on foreground regions. \textbf{Bold} numbers indicate the best quantitative results, and the \underline{underlined} numbers indicate the second best results.}
    \scalebox{0.95}{
    \setlength{\tabcolsep}{4.5pt}
    \begin{tabular}{cccccc}
    \toprule
     Data   &  AF & Method & NMSE & PSNR & SSIM\\
     \midrule
     \multirow{8}{*}{CORPD} & \multirow{4}{*}{4} & Unet & 0.0061 $\pm$ 0.0046 & 37.22 $\pm$ 3.92 & 0.9346 $\pm$ 0.0354 \\
                                                       && SN  & \textbf{0.0026 $\pm$ 0.0025} & \textbf{41.15 $\pm$ 4.50} & \textbf{0.9633 $\pm$ 0.0276} \\
                                                       && SN-GAN & 0.0028 $\pm$ 0.0026 & 40.78 $\pm$ 4.42 & 0.9605 $\pm$ 0.0287                                \\                                           && interpSN & \textbf{0.0026 $\pm$ 0.0025} & \underline{41.07 $\pm$ 4.50} & \underline{0.9629 $\pm$ 0.0278}
                                                       \\\cmidrule{3-6}
                                   & \multirow{4}{*}{8} & Unet & 0.0174 $\pm$ 0.0113 & 32.62 $\pm$ 3.49 & 0.8831 $\pm$ 0.0521 \\
                                                       && SN & \textbf{0.0062 $\pm$ 0.0051} & \textbf{37.22 $\pm$ 3.91} & \textbf{0.9327 $\pm$ 0.0405} \\
                                                       && SN-GAN & 0.0071 $\pm$ 0.0057 & 36.61 $\pm$ 3.86 & 0.9247 $\pm$ 0.0437\\
                                                       && interpSN & \underline{0.0064 $\pm$ 0.0052} & \underline{37.09 $\pm$ 3.92} & \underline{0.9313 $\pm$ 0.0410}\\
     \midrule
     \multirow{8}{*}{CORPDFS} & \multirow{4}{*}{4} & Unet & 0.0133 $\pm$ 0.0082 & 36.67 $\pm$ 3.75 & 0.9000 $\pm$ 0.0571 \\
                                                       && SN  & \textbf{0.0096 $\pm$ 0.0072} & \underline{38.18 $\pm$ 4.45} & \underline{0.9180 $\pm$ 0.0542} \\
                                                       && SN-GAN & \textbf{0.0096 $\pm$ 0.0072} & 38.16 $\pm$ 4.42 & 0.9178 $\pm$ 0.0538                                \\                                           && interpSN & \textbf{0.0096 $\pm$ 0.0072} & \textbf{38.19 $\pm$ 4.45} & \textbf{0.9182 $\pm$ 0.0540}
                                                       \\\cmidrule{3-6}
                                   & \multirow{4}{*}{8} & Unet & 0.0274 $\pm$ 0.0147 & 33.46 $\pm$ 3.72 & 0.8507 $\pm$ 0.0722  \\
                                                       && SN & \underline{0.0167 $\pm$ 0.0106} & \underline{35.67 $\pm$ 4.02} & \underline{0.8821 $\pm$ 0.0671} \\
                                                       && SN-GAN & 0.0170 $\pm$ 0.0107 & 35.58 $\pm$ 4.03 & 0.8808 $\pm$ 0.0670\\
                                                       && interpSN & \textbf{0.0167 $\pm$ 0.0105} & \textbf{35.68 $\pm$ 4.03} & \textbf{0.8823 $\pm$ 0.0669}\\
    \midrule
    \multirow{8}{*}{ALL} & \multirow{4}{*}{4} & Unet &  0.0097 $\pm$ 0.0097 & 36.95 $\pm$ 3.87 & 0.9174 $\pm$ 0.0586\\
                                                       && SN  & \underline{0.0061 $\pm$ 0.0089} & \textbf{39.67 $\pm$ 5.37} & \textbf{0.9408 $\pm$ 0.0625} \\
                                                       && SN-GAN & 0.0062 $\pm$ 0.0087 & 39.47 $\pm$ 5.14 & 0.9393 $\pm$ 0.0607                                \\                                           && interpSN & \textbf{0.0061 $\pm$ 0.0088} & \underline{39.64 $\pm$ 5.32} & \underline{0.9406 $\pm$ 0.0619}
                                                       \\\cmidrule{3-6}
                                   & \multirow{4}{*}{8} & Unet & 0.0224 $\pm$ 0.0165 & 33.04 $\pm$ 3.69 & 0.8670 $\pm$ 0.0707 \\
                                                       && SN & \textbf{0.0114 $\pm$ 0.0134} & \textbf{36.45 $\pm$ 4.25} & \textbf{0.9075 $\pm$ 0.0750} \\
                                                       && SN-GAN & 0.0121 $\pm$ 0.0131 & 36.10 $\pm$ 4.07 & 0.9029 $\pm$ 0.0715\\
                                                       && interpSN & \underline{0.0115 $\pm$ 0.0132} & \underline{36.39 $\pm$ 4.21} & \underline{0.9069 $\pm$ 0.0739} \\
    \bottomrule
\end{tabular} \label{table1}
}
\end{table}

\section{Discussion and Conclusion}
In this work, we proposed to employ a simple deep network interpolation strategy for parallel MR image reconstruction. By interpolating networks in parameter space, we showed that the new interpolated model can balance between the quantitative scores and visual perception. It is worth noting that such interpolation scheme is at no cost, and the network architecture is flexible as long as the models to be interpolated share the same structure. By varying the interpolation coefficient, we can have a smooth control of the reconstruction effects, which could potentially enable the human observer to interpret based on the adjustment between data fidelity and perceptual quality and ensure correct diagnosis. Future work can investigate on learning the interpolation coefficients to automatically find the optimal balance.

\section*{Acknowledgements}
The work was funded in part by the EPSRC Programme Grant (EP/P001009/1) and by the Intramural Research Programs of the National Institutes of Health Clinical Center.
\bibliographystyle{splncs04}
\bibliography{ref}

\begin{thebibliography}{10}
\providecommand{\url}[1]{\texttt{#1}}
\providecommand{\urlprefix}{URL }
\providecommand{\doi}[1]{https://doi.org/#1}

\bibitem{aggarwal2018modl}
Aggarwal, H.K., Mani, M.P., Jacob, M.: {MoDL: Model-based deep learning
  architecture for inverse problems}. IEEE transactions on medical imaging
  \textbf{38}(2),  394--405 (2018)

\bibitem{duan2019vs}
Duan, J., Schlemper, J., Qin, C., Ouyang, C., Bai, W., Biffi, C., Bello, G.,
  Statton, B., O’Regan, D.P., Rueckert, D.: {VS-Net: Variable splitting
  network for accelerated parallel MRI reconstruction}. In: International
  Conference on Medical Image Computing and Computer-Assisted Intervention. pp.
  713--722. Springer (2019)

\bibitem{hammernik2018learning}
Hammernik, K., Klatzer, T., Kobler, E., Recht, M.P., Sodickson, D.K., Pock, T.,
  Knoll, F.: Learning a variational network for reconstruction of accelerated
  {MRI} data. Magnetic resonance in medicine  \textbf{79}(6),  3055--3071
  (2018)

\bibitem{qin2019convolutional}
Qin, C., Schlemper, J., Caballero, J., et~al.: Convolutional recurrent neural
  networks for dynamic {MR} image reconstruction. IEEE transactions on medical
  imaging  \textbf{38}(1),  280--290 (2019)

\bibitem{qin2019k}
Qin, C., Schlemper, J., Duan, J., Seegoolam, G., Price, A., Hajnal, J.,
  Rueckert, D.: k-t {NEXT}: Dynamic {MR} image reconstruction exploiting
  spatio-temporal correlations. In: International Conference on Medical Image
  Computing and Computer-Assisted Intervention. pp. 505--513. Springer (2019)

\bibitem{schlemper2019data}
Schlemper, J., Duan, J., Ouyang, C., Qin, C., Caballero, J., Hajnal, J.V.,
  Rueckert, D.: Data consistency networks for (calibration-less) accelerated
  parallel {MR} image reconstruction. In: In ISMRM 27th Annual Meeting. p.~4664
  (2019)

\bibitem{uecker2014espirit}
Uecker, M., Lai, P., Murphy, M.J., Virtue, P., Elad, M., Pauly, J.M.,
  Vasanawala, S.S., Lustig, M.: {ESPIRiT—an eigenvalue approach to
  autocalibrating parallel MRI: where SENSE meets GRAPPA}. Magnetic resonance
  in medicine  \textbf{71}(3),  990--1001 (2014)

\bibitem{wang2019deep}
Wang, X., Yu, K., Dong, C., Tang, X., Loy, C.C.: Deep network interpolation for
  continuous imagery effect transition. In: Proceedings of the IEEE Conference
  on Computer Vision and Pattern Recognition. pp. 1692--1701 (2019)

\bibitem{yu2019deep}
Yu, S., Park, B., Jeong, J.: Deep iterative down-up {CNN} for image denoising.
  In: Proceedings of the IEEE Conference on Computer Vision and Pattern
  Recognition Workshops (2019)

\bibitem{zbontar2018fastmri}
Zbontar, J., Knoll, F., Sriram, A., Muckley, M.J., Bruno, M., Defazio, A.,
  Parente, M., Geras, K.J., Katsnelson, J., Chandarana, H., et~al.: {fastMRI:
  An open dataset and benchmarks for accelerated MRI}. arXiv preprint
  arXiv:1811.08839  (2018)

\end{thebibliography}

\end{document}